\documentclass[doublecol]{epl2} 

\usepackage{url}
\usepackage{bm,color}
\usepackage{graphics}
\usepackage{amssymb,amsfonts,amsmath}

\usepackage{pdfpages}

\title{Controlling the shape of membrane protein polyhedra}
\shorttitle{Controlling the shape of membrane protein polyhedra} 

\author{Di Li, Osman Kahraman and Christoph A. Haselwandter}
\shortauthor{D. Li \etal}

\institute{Department of Physics \& Astronomy and Molecular and Computational Biology Program, \\Department of Biological Sciences, University of Southern California, Los Angeles, CA 90089, USA
}


\abstract{
Membrane proteins and lipids can self-assemble into membrane protein polyhedral nanoparticles (MPPNs). MPPNs have a closed spherical surface and a polyhedral protein arrangement, and may offer a new route for structure determination of membrane proteins and targeted drug delivery. We develop here a general analytic model of how MPPN self-assembly depends on bilayer-protein interactions and lipid bilayer mechanical properties. We find that the bilayer-protein hydrophobic thickness mismatch is a key molecular control parameter for MPPN shape that can be used to bias MPPN self-assembly towards highly symmetric and uniform MPPN shapes. Our results suggest strategies for optimizing MPPN shape for structural studies of membrane proteins and targeted drug
delivery.
}

\begin{document}

\maketitle

\section{Introduction}

In recent experiments \cite{Basta2014}, membrane proteins and lipids were observed to self-assemble in an aqueous environment into membrane protein polyhedral nanoparticles (MPPNs)---closed lipid bilayer vesicles with a polyhedral arrangement of membrane proteins. In particular, the mechanonsensitive channel of small conductance (MscS) \cite{Bass2002,Steinbacher2007} was observed \cite{Basta2014,Wu2013} to predominantly yield MPPNs with the symmetry of a snub cube, with one MscS located at each of its 24 vertices, and a characteristic overall radius $\approx 20$~nm. Through their well-defined symmetry and characteristic size, MPPNs may \cite{Basta2014}, in addition to potential applications as novel drug delivery carriers, offer a new route for structure determination of membrane proteins, with the membrane proteins embedded in a lipid bilayer environment and the closed surfaces of MPPNs supporting physiologically relevant transmembrane gradients. We have shown previously \cite{Li2016} that the observed symmetry and size of MPPNs \cite{Basta2014} can be understood based on the interplay of protein-induced lipid bilayer curvature deformations \cite{Gozdz2001,Auth2009,Muller2010} arising \cite{Phillips2009} from the conical shape of MscS \cite{Bass2002,Steinbacher2007}, topological defects in protein packing necessitated by the spherical shape of MPPNs \cite{Bruinsma2003}, and thermal fluctuations in MPPN self-assembly \cite{Bruinsma2003,Safranbook,Benshaul1994}.

Realization of MPPNs as a novel method for membrane protein structural analysis,
as well as targeted drug delivery, requires \cite{Basta2014} control over MPPN symmetry and size. Current experimental approaches, however, yield a distribution of different MPPN shapes \cite{Basta2014,Li2016}, which limits the resolution of MPPN-based structural studies \cite{Basta2014} and potential applications of MPPNs as novel drug delivery carriers. To explore strategies for controlling and optimizing MPPN shape, we generalize here our previous model of MPPN self-assembly \cite{Li2016} to account for the effects of a hydrophobic thickness mismatch between membrane protein and the unperturbed lipid bilayer. We provide general analytic solutions for the dependence of the MPPN energy on bilayer-protein hydrophobic thickness mismatch and, on this basis, calculate a generalized \textcolor{black}{MPPN self-assembly diagram}. Our results suggest that, in addition to the bilayer-protein contact angle \cite{Li2016}, the bilayer-protein hydrophobic thickness mismatch is a key molecular control parameter for MPPN shape. In particular, we find that modification of the lipid bilayer composition, or protein hydrophobic thickness, so as to produce pronounced protein-induced lipid bilayer thickness deformations biases the \textcolor{black}{MPPN self-assembly diagram} towards highly symmetric and uniform MPPN shapes. Our results provide general insights into the roles of bilayer-protein interactions and lipid bilayer mechanical properties in MPPN self-assembly, and suggest strategies for controlling MPPN shape in experiments.

\section{Bilayer mechanics of MPPNs}

\begin{figure}[t!]
\centerline{\includegraphics[width=0.9\columnwidth]{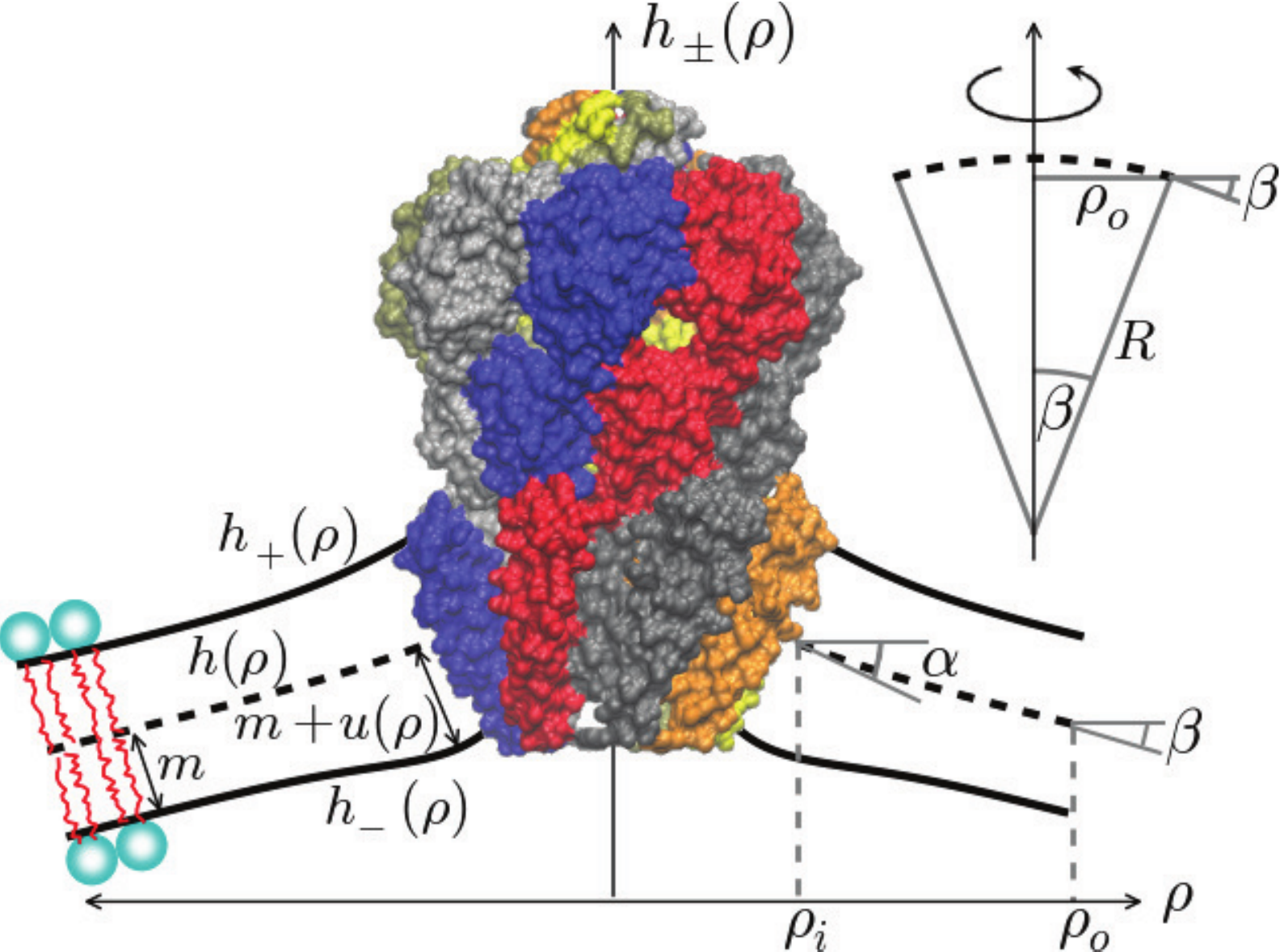}}
\caption{(Color online) Schematic of protein-induced lipid bilayer deformations
in MPPNs. We denote the bilayer midplane radius of the protein by $\rho_i$
and the bilayer-protein contact angle by $\alpha$. The transmembrane surface
of the protein specifies boundary conditions on $h$ and $u$ at the bilayer-protein
interface (see main text). \textcolor{black}{The membrane patch radius $\rho_o=R
\sin \beta$, where $R$ is the MPPN bilayer midplane radius at $\rho=\rho_o$ and the membrane patch angle $\beta=\arccos[(n-2)/n]$ is determined via the relation $4 \pi R^2=n \Omega R^2 $, in which $\Omega=2\pi (1-\cos \beta)$ is the solid angle subtended by each (circular) membrane patch and $n$ is the number of proteins per MPPN \cite{Auth2009,Li2016}
(see inset).} The protein structure shown here corresponds to the closed state of MscS \cite{Bass2002,Steinbacher2007} used in experiments on MPPNs \cite{Basta2014,Wu2013}, with Protein Data Bank ID 2OAU and different colors indicating different MscS subunits \cite{VMD1996}.}
\label{fig.cartoon}
\end{figure}

Membrane proteins are generally found to be rigid compared to lipid bilayer membranes, resulting in protein-induced lipid bilayer deformations 
\cite{Phillips2009,Andersen2007,Jensen2004,Lundbak2006}. In the standard elasticity theory of lipid bilayers 
\cite{Canham1970,Evans1974,Helfrich1973,Huang1986,Dan1993,Phillips2009,Fournier1999,Wiggins2005}, bilayer-protein interactions are captured by two coupled scalar fields $h_+(\rho)$ and $h_-(\rho)$ (see fig.~\ref{fig.cartoon}) that specify the positions of the hydrophilic-hydrophobic interface in the outer and inner lipid bilayer leaflets, respectively. For future convenience, we use here the Monge representation of $h_\pm$ and assume rotational symmetry about the protein center, with $\rho$ denoting the radial distance from the protein center. It is mathematically
convenient \cite{Bitbol2012,Fournier1999,Wiggins2005} to recast $h_\pm(\rho)$ in terms of the lipid bilayer midplane deformation field
\begin{equation}
h(\rho)=\frac{1}{2}\left[h_+(\rho)+h_-(\rho) \right]
\end{equation}
and, to leading order \cite{Fournier1999,Bitbol2012}, the lipid bilayer thickness deformation field
\begin{equation}
u(\rho)=\frac{1}{2}\left[h_+(\rho)-h_-(\rho)-2m \right]\,,
\end{equation}
where $2m$ denotes the hydrophobic thickness of the unperturbed lipid bilayer
(fig.~\ref{fig.cartoon}). The resulting elastic energies of lipid bilayer midplane deformations, $G_h$, and lipid bilayer thickness deformations, $G_u$, decouple from each other to leading order \cite{Fournier1999,Wiggins2005,Kahraman2016bilayer}:
\begin{align}
\label{eq.2}
& G_h= \int \frac{\upd A}{2} {\textstyle\left[K_b (\nabla^2 h)^2+\tau (\nabla h)^2\right]}\,,\\
& G_u=\int \frac{\upd A}{2}{\textstyle \left\{K_b (\nabla^2 u)^2+K_t\left( \frac{u}{m}\right)^2+\tau
\left[2 \frac{u}{m}+(\nabla u)^2\right]\right\}}\,,\label{eq.3}
\end{align}
where d$A=2 \pi \rho\,\upd\rho$, $K_b$ is the lipid bilayer bending rigidity, $\tau$ is the membrane tension, and $K_t$ is the stiffness associated with lipid bilayer thickness deformations. For generality we consider in eq.~(\ref{eq.3})
the term $2\tau u/m$, which accounts for stretching deformations tangential
to the leaflet surfaces \cite{Safranbook,Ursell2007,Ursell2008}, as well as the term $\tau (\nabla u)^2$, which accounts for changes in the projection of the bilayer area onto the reference plane used in the Monge representation of $h_\pm$ \cite{Boal2012book,Wiggins2004,Wiggins2005}. \textcolor{black}{During MPPN self-assembly \cite{Basta2014,Wu2013}, MPPNs are not expected to be able to support transmembrane gradients, suggesting that $\tau=0$. For completeness, however, we provide, below, general analytic expressions of $h$ and $u$ for arbitrary $\tau$. These analytic expressions of $h$ and $u$ could be used, for instance, to determine the shape of MPPNs if a finite $\tau$ is induced after MPPN self-assembly is completed \cite{Basta2014}.} For the diC14:0 lipids \cite{dic14} used for MPPNs formed from MscS \cite{Basta2014,Wu2013}, we have \cite{Rawicz2000,Wiggins2005} $K_b\approx 14$~$k_B T$, $K_t\approx 56.5$~$k_B T/\text{nm}^2$, and $m\approx 1.76$~nm. Unless indicated otherwise we use, throughout this letter, the parameter values associated with MPPNs formed from MscS \cite{Basta2014,Wu2013} for numerical calculations.

The protein-induced lipid bilayer deformations in eqs.~(\ref{eq.2}) and~(\ref{eq.3})
yield \cite{Goulian1993,Dan1993,Weikl1998,Phillips2009,Haselwandter2013,Kahraman2016bilayer} bilayer-mediated interactions between membrane proteins in MPPNs. For the case of rotationally-symmetric membrane inclusions considered here, $G_h$ and $G_u$ are both expected to favor hexagonal protein arrangements 
\cite{Gozdz2001,Auth2009,Muller2010,Dan1994,Fournier1999,Weitz2013,Kahraman2016architecture}.
\textcolor{black}{The resulting contributions to the MPPN energy can be calculated \cite{Li2016} from eqs.~(\ref{eq.2}) and~(\ref{eq.3}), at the mean-field level, by approximating the hexagonal unit cell by a circular membrane patch \cite{Gozdz2001,Auth2009,Muller2010,Dan1994,Fournier1999}
of radius $\rho_o$ (fig.~\ref{fig.cartoon}). The membrane patch radius $\rho_o$ depends on the number of proteins per MPPN, $n$, and the
MPPN bilayer midplane radius at the outer membrane patch boundary, $R$, via $\rho_o=R \sin \beta$ (see inset in fig.~\ref{fig.cartoon}), where the membrane patch angle $\beta =\arccos[(n-2)/n]$ \cite{Li2016,Auth2009}.} Through minimization of $G_h$ and $G_u$ in eqs.~(\ref{eq.2}) and~(\ref{eq.3}) with respect to $h(\rho)$ and $u(\rho)$ in each membrane patch we derive, below, general analytic expressions for the MPPN midplane and thickness deformation energies, $E_h(n,R)$ and $E_u(n,R)$ [see eqs.~(\ref{eq.8}) and~(\ref{eq.23})].

The spherical shape of MPPNs necessitates topological defects in the preferred hexagonal packing of membrane proteins which, in analogy to viral capsids \cite{Bruinsma2003,Zandi2004}, yields \cite{Li2016} an energy penalty characteristic of $n$. This energy penalty can be quantified \cite{Li2016,Bruinsma2003}, at the mean-field level, by approximating the spring network associated with the preferred hexagonal protein arrangements \cite{Auth2009} by a uniform elastic sheet \cite{Kantor1987,Phillips2012} with stretching modulus
\begin{equation} \label{Ksdef}
K_s= \frac{\sqrt{3}}{24\,n} \frac{\partial^2 E_0}{\partial {\rho_o}^2}\bigg|_{\rho_o=\rho_\text{min}}\,,
\end{equation}
where $E_0=E_h+E_u$ and $\rho_\text{min}\geq\rho_i$ corresponds to the minimum of $E_0$ yielding the lowest MPPN energy, in which $\rho_i$ is the protein radius in the lipid bilayer midplane with $\rho_i\approx 3.2$~nm for MscS \cite{Bass2002,Steinbacher2007,Li2016} (fig.~\ref{fig.cartoon}). We quantify, at the mean-field level, the deviation from the preferred hexagonal packing of membrane proteins due to the spherical shape of MPPNs through \cite{Bruinsma2003,Li2016} the fraction of the surface of a sphere enclosed by $n$ identical non-overlapping circles at closest packing \cite{Clare1991}, $p(n)$, resulting in the MPPN defect energy \cite{Li2016}
\begin{equation}
E_d(n,R)=2 \pi K_s R^2 \left[\frac{p_{\text{max}}-p(n)}{p_{\text{max}}}\right]^2\,,
\label{eq.9}
\end{equation}
where $p_{\text{max}}=\pi/2\sqrt{3}$ corresponds to uniform hexagonal protein arrangements. We calculate the MPPN energy $E_\text{min}(n)$ by minimizing the sum of $E_h$, $E_u$, and $E_d$ at each $n$ with respect to $R$. To account for steric constraints on lipid and protein size we only allow \cite{Li2016} for membrane patch sizes $\geqslant \rho_i+\rho_l$ when calculating $E_\text{min}(n)$, where the lipid radius $\rho_l\approx 0.45$~nm for the diC14:0 lipids \cite{dic14,Damodaran1993} used for MPPNs formed from MscS \cite{Basta2014,Wu2013}.

\subsection{MPPN midplane deformation energy}

The Euler-Lagrange equation associated with $G_h$ in eq.~(\ref{eq.2}) is given by
$\Delta^2 h = \xi^2 \Delta h$, where $\xi=\sqrt{\tau/K_b}$ is the inverse
decay length of midplane deformations \cite{Wiggins2005}, with the general
solution \cite{Weikl1998}
\begin{equation}
h(\rho) = A_h I_0(\xi\rho)+B_h K_0(\xi\rho)+C_h+D_h \ln\rho\,,
\label{eq.6}
\end{equation}
where $I_0$ and $K_0$ are the zeroth-order modified Bessel functions of the first and second kind, respectively. The constants $A_h$, $B_h$, $C_h$,
and $D_h$ in eq.~(\ref{eq.6}) are determined by the boundary conditions along
the bilayer-protein interface and the outer boundary of the membrane patch. In particular, the slope of the lipid bilayer at the bilayer-protein interface is given by $h'(\rho_i) \equiv a = -\tan\alpha$, with the bilayer-protein contact angle $\alpha \approx 0.46$--$0.54$~rad for MscS \cite{Bass2002,Steinbacher2007,Li2016}
(fig.~\ref{fig.cartoon}). The slope at the outer boundary of the membrane
patch is given by $h'(\rho_o) \equiv b = -\tan\beta$ \cite{Li2016,Auth2009}, which enforces the spherical shape of MPPNs (fig.~\ref{fig.cartoon}). Furthermore, we impose \cite{Li2016,Auth2009} a zero-force boundary condition \cite{Weikl1998} at $\rho=\rho_o$,
\begin{equation} \label{zeroFh}
\frac{\partial}{\partial\rho}\left[\Delta h(\rho)-\xi^2 h(\rho) \right]\bigg|_{\rho=\rho_o}=0\,,
\end{equation}
and fix the (arbitrary) reference point of $h$ via $h(\rho_i)=0$. These four boundary conditions, together with eq.~(\ref{eq.6}), imply that
\begin{eqnarray}
A_h&=&\frac{b K_1(\xi \rho_i)-a K_1(\xi \rho_o)}{F}\,, \label{eq.ct1} \\
B_h&=&\frac{b I_1(\xi \rho_i)-a I_1(\xi \rho_o)}{F}\,, \label{eq.ct2} \\
C_h&=&\frac{a K_0(\xi \rho_i) I_1(\xi \rho_o)+a I_0(\xi \rho_i) K_1(\xi \rho_o)-b/(\xi \rho_i)}{F}\,,\nonumber\\&& \label{eq.ct3}
\end{eqnarray}
and $D_h=0$, where 
\begin{equation}
F= \xi [K_1(\xi \rho_i) I_1(\xi \rho_o)-I_1(\xi \rho_i) K_1(\xi \rho_o)]\,,
\end{equation}
and $I_1$ and $K_1$ are the first-order modified Bessel functions of the first and second kind, respectively. Integration of eq.~(\ref{eq.2}) with
eq.~(\ref{eq.6}) from $\rho=\rho_i$ to $\rho=\rho_o$ for all $n$ membrane patches thus results
in the MPPN midplane deformation energy
\begin{align} \nonumber
E_h(n,R) = n \pi \tau \big\{&b \rho_o \left[A_h I_0(\xi \rho_o)+B_h K_0(\xi \rho_o)\right] \\
&-a \rho_i \left[A_h I_0(\xi \rho_i)+B_h K_0(\xi \rho_i)\right]\big\}\,.
\label{eq.8}
\end{align}
For $\rho_o\rightarrow \infty$ and $b\rightarrow 0$ \cite{Auth2009}, eq.~(\ref{eq.8}) yields the minimum of the midplane deformation energy in eq.~(\ref{eq.2}) for a single conical inclusion ($n=1$) in an infinite, asymptotically flat lipid bilayer membrane \cite{Weikl1998,Wiggins2005}, while for $\tau\to 0$ we recover the results in refs.~\cite{Auth2009,Li2016}.

\subsection{MPPN thickness deformation energy}

The Euler-Lagrange equation associated with $G_u$ in eq.~(\ref{eq.3}) is given~by
\begin{equation}
(\Delta - \nu_+)(\Delta - \nu_-)\bar{u} = 0\,,
\label{eq.19}
\end{equation}
where $\bar{u}(\rho) = u(\rho) + \frac{\tau m}{K_t}$ and
\begin{equation}
\nu_{\pm}=\frac{1}{2K_b}\left(\tau \pm \sqrt{\tau^2-\frac{4K_b K_t}{m^2}}\right)\,.
\label{eq.18}
\end{equation}
Equation~(\ref{eq.19}) has the solution \cite{Huang1986,Zauderer,Kahraman2016bilayer}
\begin{align}
&\bar{u}(\rho) = A^+_u K_0(\sqrt{\nu_+}\rho)+A^-_u K_0(\sqrt{\nu_-}\rho) \nonumber \\
& \qquad \quad + B^+_u I_0(\sqrt{\nu_+}\rho)+B^-_u I_0(\sqrt{\nu_-}\rho)\,,
\label{eq.mgc}
\end{align}
where the constants $A^\pm_u$ and $B^\pm_u$ are fixed by the boundary conditions
on $u(\rho)$ along the bilayer-protein interface and the outer boundary of the membrane patch, respectively.

To determine the boundary conditions on $u(\rho)$ we first note \cite{Dan1993,Dan1994} that, by symmetry, $u'(\rho_o)=0$ in our mean-field model of MPPNs. We also set $u'(\rho_i)=0$, which is consistent with experiments on gramicidin channels \cite{Huang1986,Harroun1999,Harroun1999b} and the mechanosensitive channel of large conductance \cite{Wiggins2004,Wiggins2005,Ursell2007,Ursell2008},
but other choices for this boundary condition could also be implemented 
\cite{Dan1993,Dan1994,Nielsen1998,Partenskii2002,Partenskii2003,Partenskii2004,Brannigan2006,Brannigan2007,West2009,Bitbol2012,Lee2013}. Furthermore, we assume \cite{Huang1986,Dan1993,Dan1994,Andersen2007,Phillips2009}
that the lipid bilayer deforms along the bilayer-protein interface so as to match the hydrophobic thickness of the membrane protein, yielding $u(\rho_i)=U$,
with the hydrophobic thickness mismatch $U=\frac{1}{2} W - m$,
where $W\approx 3.63$~nm for MscS \cite{Bass2002,Steinbacher2007} so that $U\approx 0.055$~nm for MPPNs formed from MscS and diC14:0 lipids \cite{Basta2014,Wu2013,dic14,Rawicz2000}. Finally, a fourth boundary condition is obtained by letting $u(\rho_o)$ vary
so as to minimize the thickness deformation energy, which amounts \cite{Weikl1998,Auth2009}
to a zero-force boundary condition at $\rho=\rho_o$ analogous to eq.~(\ref{zeroFh}):
\begin{equation} \label{eqBCu}
\left.\frac{\partial}{\partial \rho}\left[\Delta \bar{u}(\rho)-\xi^2\bar{u}(\rho)\right]\right|_{\rho=\rho_o}=0\,,
\end{equation}
where, as in eq.~(\ref{zeroFh}), $\xi=\sqrt{\tau/K_b}$. Note that the zero-force boundary condition in eq.~(\ref{eqBCu}) does not explicitly depend on the terms in eq.~(\ref{eq.3}) that only involve $u$, and not its derivatives.

Together with eq.~(\ref{eq.mgc}), the above boundary conditions imply that
\begin{equation}
A^\pm_{u} = \frac{\pm I^\pm_{1o} (Q^{\mp \mp \mp}_{1o1i}-Q^{\mp \mp \mp}_{1i1o})
\left(U+\frac{\tau m}{K_t}\right)}{S}\,,
\label{eq.cu1}
\end{equation}
\begin{equation}
B^\pm_{u} = \frac{\pm K^\pm_{1o} (Q^{\mp \mp \mp}_{1o1i}-Q^{\mp \mp \mp}_{1i1o})
\left(U+\frac{\tau m}{K_t}\right)}{S}\,,
\label{eq.cu2}
\end{equation}
where
\begin{align}
&S = I^-_{1o} \left[(P^{- + -}_{0i1i}-P^{+ - +}_{0i1i}) I^+_{1o}+(Q^{++-}_{1i0i}+Q^{-+-}_{0i1i}) K^+_{1o}\right] \nonumber\\
&\qquad + K^-_{1o} \left[(Q^{+++}_{1i1o}-Q^{+++}_{1o1i}) I^-_{0i}-(Q^{-++}_{1o0i}+Q^{-++}_{0i1o}) I^-_{1i}\right]\,,
\label{eq.15}
\end{align}
and we define, for $j=0,1$, $l=0,1$, $\eta=i,o$, and $\theta=i,o$,
\begin{eqnarray}
P^{\pm\pm\pm}_{j\eta l \theta} &\equiv& \sqrt{\nu_{\pm}}K^{\pm}_{j \eta}K^{\pm}_{l
\theta}\,,\\
Q^{\pm\pm\pm}_{j \eta l \theta} &\equiv& \sqrt{\nu_{\pm}}I^{\pm}_{j \eta}K^{\pm}_{l \theta}\,,
\end{eqnarray}
and $K^{\pm}_{j\eta} \equiv K_j(\sqrt{\nu_{\pm}}\rho_\eta)$ and $I^{\pm}_{j\eta} \equiv I_j(\sqrt{\nu_{\pm}}\rho_\eta)$. Integration of eq.~(\ref{eq.3}) with eq.~(\ref{eq.mgc}) from $\rho=\rho_i$ to $\rho=\rho_o$ for all $n$ membrane
patches thus results in the MPPN thickness deformation energy
\begin{align}
&E_u(n,R) = n \pi\bigg\{ K_b \rho_i \left(U+\frac{\tau m}{K_t}\right) \bigg[\nu_+^{3/2} \left(B^+_u I^+_{1i}-A^+_u K^+_{1i}\right) \nonumber\\
& \qquad \qquad + \nu_-^{3/2} (B^-_u I^-_{1i}-A^-_u K^-_{1i})\bigg] + \frac{\tau^2}{2 K_t}\left(\rho_i^2 - \rho_o^2\right) \bigg\}\,.
\label{eq.23}
\end{align}
For $\rho_o\rightarrow \infty$, eq.~(\ref{eq.23}) reproduces previous results \cite{Huang1986,Wiggins2005,Haselwandter2013connection} on the minimum of the thickness deformation energy in eq.~(\ref{eq.3}) for a single cylindrical inclusion ($n=1$) in an infinite, asymptotically flat lipid bilayer membrane.

\section{MPPN energy}

\begin{figure}
\centerline{\includegraphics[width=0.96\columnwidth]{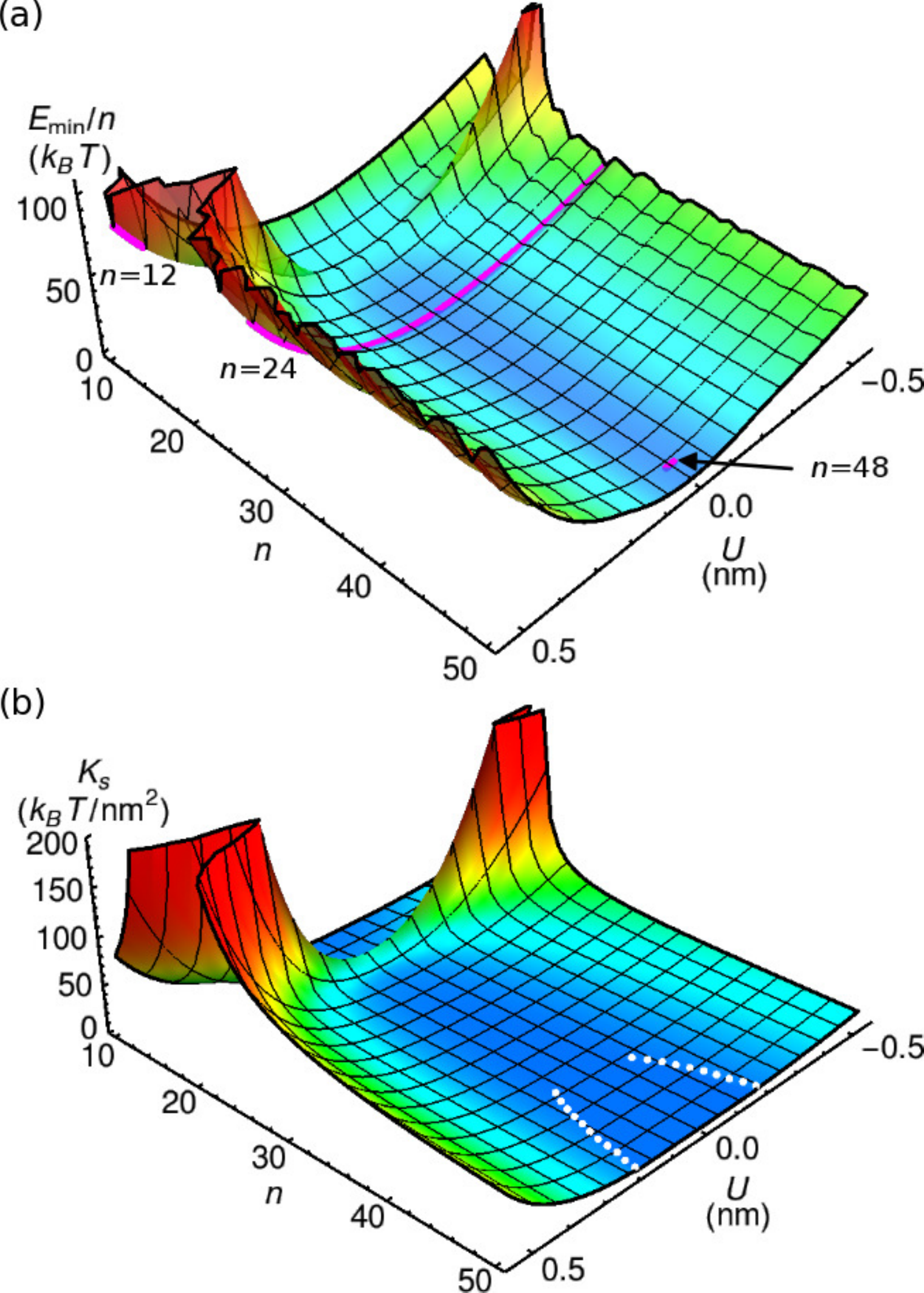}}
\caption{(Color online) MPPN energy and MPPN stretching modulus for MPPNs formed from MscS \cite{Bass2002,Steinbacher2007} at $\alpha=0.5$~rad for $\tau=0$. (a) MPPN energy per protein, $E_\text{min}/n$, obtained from eqs.~(\ref{eq.9}),~(\ref{eq.8}), and~(\ref{eq.23}) \textit{vs.} $n$ and $U$. The magenta curves show the $n$-states with minimal $E_\text{min}/n$, for $10 \leq n \leq 80$, as a function of $U$. (b) MPPN stretching modulus $K_s$ obtained from eq.~(\ref{Ksdef}) \textit{vs.} $n$ and $U$. The white dashed curves indicate locations in parameter space with a discontinuous jump in $K_s$ as a function of $U$. In both panels, $U$ is changed by varying~$m$.}
\label{fig.2}
\end{figure}

\textcolor{black}{To calculate the MPPN energy $E_\text{min}(n)$ we minimize the sum of $E_d(n,R)$ in eq.~(\ref{eq.9}),~$E_h(n,R)$ in eq.~(\ref{eq.8}), and $E_u(n,R)$ in eq.~(\ref{eq.23}) at each $n$ with respect to $R$, from which we obtain the MPPN energy per protein, $E_\text{min}(n)/n$, with all remaining
model parameters (i.e., $U$, $\rho_i$, $\alpha$, $m$, $K_b$, and $K_t$) determined directly by the molecular properties of the lipids and proteins forming MPPNs [see fig.~\ref{fig.2}(a)].} \textcolor{black}{We vary $m$ to produce values of $U$ between $U\approx -0.5$~nm and $U \approx 0.5$~nm, where $U\approx 0.055$~nm
with $m\approx 1.76$~nm corresponds to \cite{Bass2002,Steinbacher2007,Rawicz2000,Wiggins2005} the diC14:0 lipids \cite{dic14} used for MPPNs formed from MscS \cite{Basta2014,Wu2013}.
Such a $U$-range could potentially be realized in experiments on MPPNs \cite{Basta2014,Wu2013} by using lipids with different acyl-chain lengths \cite{Rawicz2000,Perozo2002,Wiggins2004,Wiggins2005}.
Varying $m$ generally also modifies the values of $K_b$ and $K_t$ \cite{Rawicz2000}. For simplicity we employ, for now, the values $K_b = 14$~$k_B T$ and $K_t = 56.5$~$k_B T/\text{nm}^2$ \cite{Rawicz2000,Wiggins2005} associated with the diC14:0 lipids \cite{dic14} used for MPPNs formed from MscS \cite{Basta2014,Wu2013}.
We return, below, to the effect of variations in $K_b$ and $K_t$ with $m$.}
Note that, if $U$ is changed by varying $m$, $E_\text{min}/n$ is not invariant under $U\to-U$ because $G_u$ in eq.~(\ref{eq.3}) explicitly depends on $m$. 

We find that the magnitude of the MPPN energy tends to increase with increasing $|U|$ [fig.~\ref{fig.2}(a)], because $E_u$ increases with $|U|$ \cite{Wiggins2005}. Similarly as for the case $U=0$ \cite{Li2016}, the contribution $E_d$ to $E_\text{min}$ yields, also for $U \neq 0$, a series of local minima in $E_\text{min}/n$ at locally optimal protein packing states \cite{Clare1991,Li2016}. We find that, at small $|U|$, $n=48$ provides the minimum of $E_\text{min}/n$ in the range $10 \leq n \leq 80$, with several competing $n$ yielding $E_\text{min}/n$ within a fraction of $k_B T$ of $n=48$. As $|U|$ is increased, we find MPPNs with snub cube symmetry, $n=24$, as well as icosahedral symmetry, $n=12$, as the minima of $E_\text{min}/n$ in the range $10 \leq n \leq 80$. This can be understood by noting that bilayer-thickness-mediated interactions between integral membrane proteins favor close packing of membrane proteins \cite{Dan1994,Kahraman2016architecture}. Particularly favorable protein packing states such as the icosahedron and the snub cube therefore become dominant as $|U|$ is increased, with the icosahedron providing \cite{Clare1991} the optimal protein packing for $10 \leq n \leq 80$. Note that the transition from $n=24$ to $n=12$ as the minimum of $E_\text{min}/n$ with increasing $|U|$ only occurs for $U>0$ in fig.~\ref{fig.2}(a). This can be understood by noting that, for $U>0$, $m$ is smaller than for $U<0$, yielding a larger magnitude of $G_u$ in eq.~(\ref{eq.3}).

\begin{figure}[t!]
\centerline{\includegraphics[width=0.96\columnwidth]{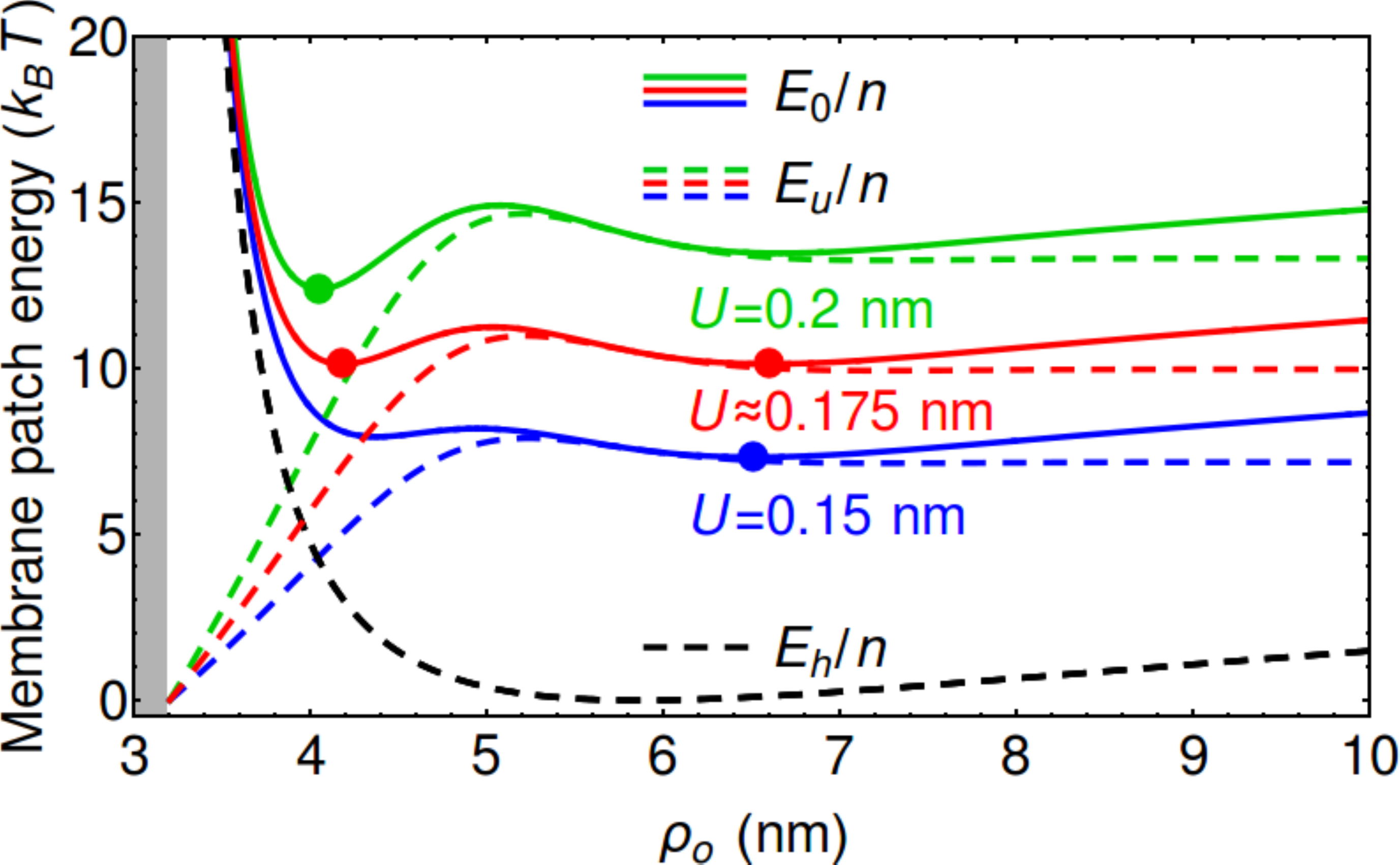}}
\caption{(Color online) Bilayer midplane and thickness deformation energies per membrane patch, $E_h/n$ and $E_u/n$, obtained from eqs.~(\ref{eq.8}) and~(\ref{eq.23}), and $E_0/n=(E_h+E_u)/n$ \textit{vs.}~$\rho_o$ for MPPNs formed from MscS \cite{Bass2002,Steinbacher2007} at $\alpha=0.5$~rad for $\tau=0$, $n=48$, and the indicated values of $U$, which we obtain by varying
$m$. The values of $\rho_o$ yielding global minima of $E_0/n$ for $\rho_o \geq \rho_i$ are indicated by dots, with a two-fold degenerate minimum of $E_0/n$ at $U\approx 0.175$~nm. The grey shaded region indicates the range $\rho_o < \rho_i$ excluded by steric constraints.}
\label{fig.ks}
\end{figure}

The MPPN stretching modulus $K_s$ entering the MPPN defect energy in eq.~(\ref{eq.9}) tends to increase with increasing $|U|$ [see fig.~\ref{fig.2}(b)].
Similarly as for $E_\text{min}$, $K_s$ is not invariant under $U\to-U$ if,
as in fig.~\ref{fig.2}, $U$ is changed by varying $m$, because $G_u$ in eq.~(\ref{eq.3}) explicitly depends on $m$. $K_s$ takes particularly large values at $n\approx 16$ in fig.~\ref{fig.2}(b) because \cite{Li2016} the membrane patch radius $\rho_o=\rho_\textrm{min}$ in eq.~(\ref{Ksdef}) approaches $\rho_i$ for $n \approx 16$. The continuum model of MPPN bilayer mechanics used here may not give reliable results in this regime. We also find that, for certain $n$, $K_s$ is a discontinuous function of $U$ in fig.~\ref{fig.2}(b). This can be understood by noting that $\rho_\textrm{min}$ in eq.~(\ref{Ksdef})
depends crucially on the competition between $E_h$, which yields short-range repulsion and long-range attraction \cite{Auth2009} between membrane proteins in MPPNs, and $E_u$, which favors the smallest $\rho_o$ allowed by steric constraints, but also yields a local energy minimum at intermediate $\rho_o$ \cite{Dan1993,Dan1994,Kahraman2016bilayer,Haselwandter2013directional}
(see fig.~\ref{fig.ks}). For $n<40$ with $U>0$ ($n<41$ with $U<0$) in fig.~\ref{fig.2}(b), $\rho_\textrm{min}$ always lies within the small-$\rho_o$ regime of $E_u$. But, for $n\geq 40$ with $U>0$ ($n\geq 41$ with $U<0$) and small $|U|$ in fig.~\ref{fig.2}(b), $\rho_\textrm{min}$ falls into the intermediate-$\rho_o$ regime of $E_u$. As $|U|$ is increased, the magnitude of $E_u$ increases while $E_h$ remains constant (fig.~\ref{fig.ks}). As a result, for $n\geq 40$ with $U>0$ ($n\geq 41$ with $U<0$) and large enough $|U|$ in fig.~\ref{fig.2}(b), we find a transition in the position of $\rho_\textrm{min}$, from the intermediate-$\rho_o$ regime of $E_u$ to the small-$\rho_o$ regime of $E_u$, resulting in a discontinuous jump in $\rho_\textrm{min}$ and, hence, $K_s$. The discontinuity in $K_s$ in fig.~\ref{fig.2}(b) with increasing $|U|$ is accompanied, for a given $n$, by a discontinuous decrease in the preferred
MPPN radius.

\section{\textcolor{black}{MPPN self-assembly diagram}}

To calculate the \textcolor{black}{MPPN self-assembly diagram} we note \cite{Li2016} that MPPNs were obtained \cite{Wu2013,Basta2014} in dilute, aqueous solutions with a small protein number fraction $c=\sum_n N_n/N_w \approx7.8\times10^{-8}$, where $N_n$ denotes the total number of proteins bound in MPPNs with $n$ proteins each and $N_w$ denotes the total number of solvent molecules in the system, which we take to be dominated by contributions due to water. In the dilute limit $c \ll 1$ with no interactions between MPPNs, minimization of the Helmholtz free energy of the system with respect to the MPPN number fraction $\Phi(n)=N_n/n N_w$ \cite{Safranbook,Benshaul1994,Bruinsma2003}
yields \cite{Li2016}
\begin{equation}
\Phi(n)=e^{[\mu n-E_{\text{min}}(n)]/k_B T}\ ,
\label{eq.24}
\end{equation}
where the MPPN energy $E_\textrm{min}(n)$ is determined by eqs.~(\ref{eq.9}),~(\ref{eq.8}), and~(\ref{eq.23}) as described above, and the protein chemical potential $\mu$ is fixed by the constraint $\sum_n n \Phi(n)=c$ imposing a fixed protein number fraction in the system. For simplicity, we restrict $n$ to the range $10 \leq n \leq 80$, yielding the MPPN equilibrium distribution $\phi(n)=\Phi(n)/\sum_{n=10}^{80} \Phi(n)$.

\begin{figure}[t!]
\centerline{\includegraphics[width=\columnwidth]{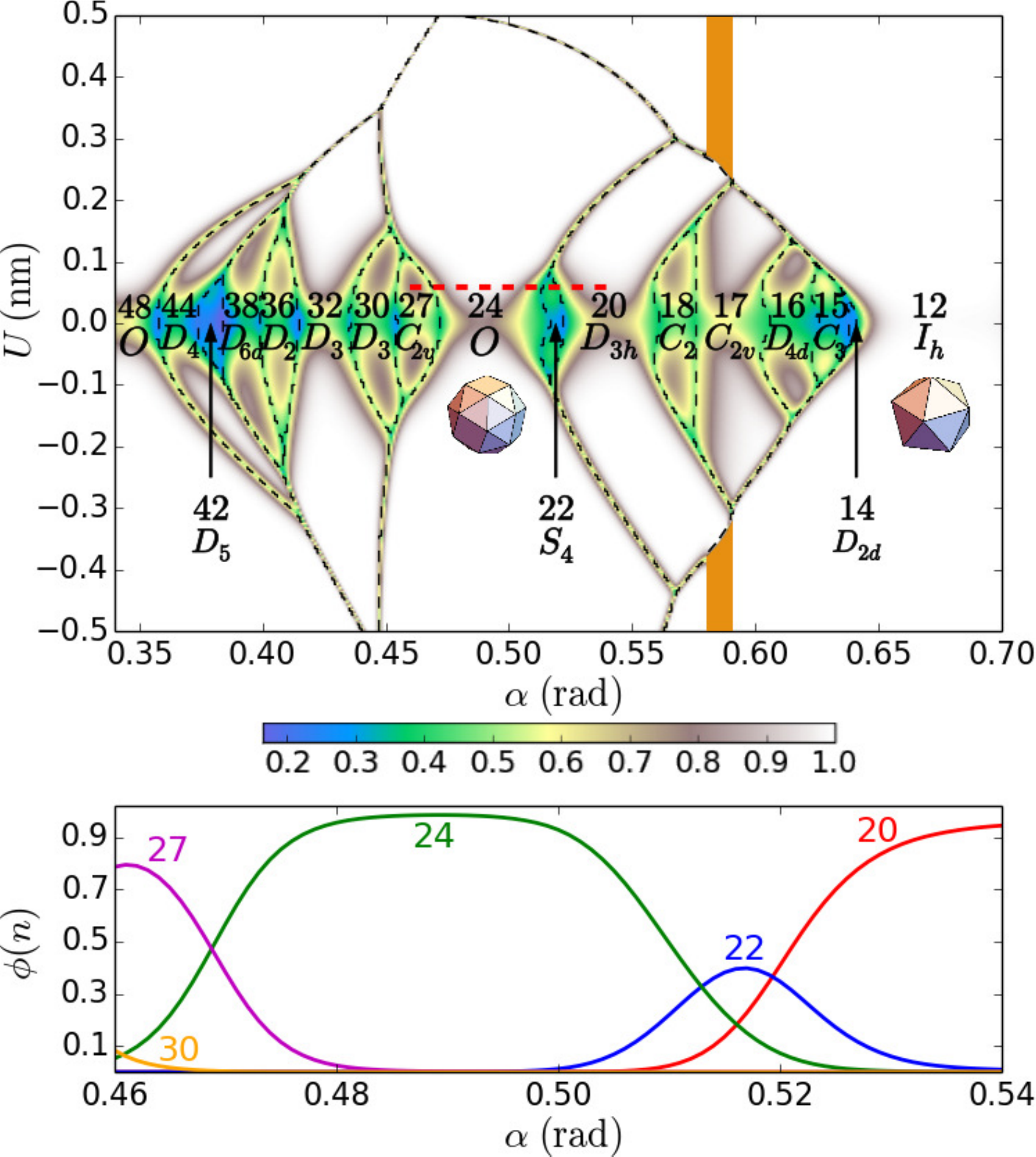}}
\caption{(Color online) \textcolor{black}{MPPN self-assembly diagram} obtained from eq.~(\ref{eq.24}) with $E_\textrm{min}(n)$ determined by eqs.~(\ref{eq.9}),~(\ref{eq.8}), and~(\ref{eq.23}) as a function of bilayer-protein hydrophobic thickness mismatch $U$, which we change by varying $m$, and bilayer-protein contact angle $\alpha$. The color map in the upper panel shows the maximum values of $\phi(n)$ associated with the dominant $n$-states of MPPNs. The dominant $n$ are indicated in each portion of the \textcolor{black}{MPPN self-assembly diagram}, together with the associated MPPN symmetry \cite{Clare1991}. Black dashed curves delineate \textcolor{black}{regions of parameter space dominated
by distinct $n$-states of MPPNs}. The red dashed horizontal line indicates the value $U\approx 0.055$~nm corresponding to the bilayer-protein hydrophobic thickness mismatch associated with \cite{Wu2013,Basta2014} MPPNs formed from MscS \cite{Bass2002,Steinbacher2007} and diC14:0 lipids \cite{dic14,Rawicz2000}, and the $\alpha$-range associated with MscS \cite{Bass2002,Steinbacher2007,Li2016}. The lower panel shows $\phi(n)$ for $n = 20$, 22, 24, 27, and 30 as a function of $\alpha$ along the red dashed horizontal line in the upper panel. We used the protein number fraction $c \approx7.8\times10^{-8}$ employed in experiments on MPPNs formed from MscS \cite{Basta2014,Wu2013}, and set $\tau=0$. The orange shaded areas indicate regions in the \textcolor{black}{MPPN self-assembly diagram} for which $n=12$ is strongly penalized by large values of $K_s$ resulting \cite{Li2016} from $\rho_o=\rho_\textrm{min}\to \rho_i$ in eq.~(\ref{Ksdef}) [see also fig.~\ref{fig.2}(b)]. In these regions of parameter space, the continuum model of MPPN bilayer mechanics used here may not give reliable results for the dominant $n$-states of MPPNs.}
\label{fig.3}
\end{figure}

Figure~\ref{fig.3} shows the \textcolor{black}{MPPN self-assembly diagram} as a function of bilayer-protein hydrophobic thickness mismatch $U$ and bilayer-protein contact angle $\alpha$ for the protein number fraction $c \approx7.8\times10^{-8}$ used in experiments on MPPNs formed from MscS \cite{Basta2014,Wu2013}. The lower panel in fig.~\ref{fig.3} provides the MPPN fractions $\phi(n)$ for the $n$-states dominant in the region of parameter space associated with \cite{Wu2013,Basta2014} MPPNs formed from MscS \cite{Bass2002,Steinbacher2007} and diC14:0 lipids \cite{dic14,Rawicz2000,Damodaran1993}, which is indicated by a dashed horizontal line in the upper panel in fig.~\ref{fig.3}. In agreement with experiments \cite{Basta2014,Wu2013} and our previous results for $U=0$ \cite{Li2016}, we find that MPPNs with snub cube symmetry, $n=24$, are dominant for MPPNs formed from MscS. Figure~\ref{fig.3} shows that, compared to the case $U = 0$ \cite{Li2016}, MscS-induced lipid bilayer thickness deformations
enhance the dominance of MPPNs with $n=24$. Apart from the dominant MPPNs with $n=24$, we also find sub-dominant MPPNs with $n=20$, $D_{3h}$ symmetry, and a MPPN radius that is reduced by $\approx 1$~nm compared to MPPNs with $n=24$. Again, these results are consistent with experiments \cite{Basta2014} as well as our previous results for $U=0$ \cite{Li2016}.

Figure~\ref{fig.3} suggests that, in addition to $\alpha$ \cite{Li2016},
the bilayer-protein hydrophobic thickness mismatch $U$ is a key molecular
parameter controlling MPPN shape. We find that, as the magnitude of $U$ is being increased and contributions due to protein-induced lipid bilayer thickness deformations come to dominate the MPPN energy, highly symmetric protein packings such as $n=12$, $n=24$, and $n=48$ \cite{Clare1991} become increasingly dominant over large portions of the \textcolor{black}{MPPN self-assembly diagram}. This can be understood by noting that bilayer-thickness-mediated interactions between integral membrane proteins favor close packing of membrane proteins \cite{Dan1994,Kahraman2016architecture}, making MPPN states with large packing fractions $p(n)$ \cite{Clare1991} strongly favorable, from an energetic perspective, for large $|U|$ [see also fig.~\ref{fig.2}(a)]. Indeed, the icosahedron, $n=12$, provides the largest value of $p(n)$ for the $n$-range considered here \cite{Clare1991}. We find that the MPPN radii $R$ of the dominant MPPNs with $n=12$, $24$, and $48$ in fig.~\ref{fig.3} only show small variations with $U$ and $\alpha$ for the parameter ranges in fig.~\ref{fig.3}. Figure~\ref{fig.3} thus suggests that protein-induced lipid bilayer thickness deformations tend to bias MPPN self-assembly towards highly symmetric and uniform MPPN shapes. In particular, we find, with all model parameters determined directly by experiments 
\cite{Phillips2009,Bass2002,Wu2013,Basta2014,Steinbacher2007,Rawicz2000,Damodaran1993}, $R\approx7$~nm, $10$~nm, $14$~nm for the regions in the \textcolor{black}{MPPN self-assembly diagram} in fig.~\ref{fig.3} for which $n=12$, $24$, and $48$ are dominant, respectively. Adjusting \cite{Li2016} for the length of the MscS cytoplasmic region $\approx 10$~nm \cite{Steinbacher2007}, the value of $R$ predicted by our model of MPPN self-assembly for $n=24$ is in quantitative agreement \cite{Li2016} with the MPPN size observed experimentally \cite{Basta2014} for $n=24$. \textcolor{black}{Finally we note that, if $U$ is varied by changing
$m$ as in fig.~\ref{fig.3}, the parameters $K_b$ and $K_t$ will generally
also vary with $U$. To check the robustness of our model predictions with respect to variations in $K_b$ and $K_t$ with $m$, we re-calculated the MPPN self-assembly diagram in fig.~\ref{fig.3} allowing for variations in $K_b$ and $K_t$ over the range of values suggested by experiments \cite{Rawicz2000}.
While we find \cite{sm} that, allowing for varying $K_b$ and $K_t$, the boundaries of regions of parameter space dominated by distinct $n$-states of MPPNs in fig.~\ref{fig.3} are shifted, the key model predictions described above remain unchanged.}

\section{Conclusion}

MPPNs constitute a novel form of ordered lipid-protein assembly intermediate between single particles and large crystalline structures \cite{Wu2013,Basta2014}.
MPPNs hold the promise of allowing structural studies of membrane proteins in the presence of physiologically relevant transmembrane gradients \cite{Wu2013,Basta2014}, and may permit \cite{Wu2013,Basta2014} targeted drug delivery with precisely controlled release mechanisms. Realization of MPPNs as a novel method for membrane protein structural analysis, and targeted drug delivery, requires \cite{Basta2014} control over MPPN shape. Our results suggest that, in addition to the bilayer-protein contact angle $\alpha$ \cite{Li2016}, the bilayer-protein hydrophobic thickness mismatch $U$ is a key molecular control parameter for MPPN shape. It has been proposed \cite{Phillips2009,Suchyna2004} that $\alpha$ can be perturbed through addition of peptide toxins that localize to the bilayer-protein interface. Our results suggest \cite{Li2016} that, in general, small effective $\alpha$ yield MPPNs with large $n$, and vice versa. However, it may be experimentally challenging to tune the effective $\alpha$ associated with a given integral membrane protein of unknown structure with sufficient precision so as to produce a particular MPPN symmetry. In contrast, a range of $U$ can be generated experimentally \cite{Phillips2009,Andersen2007,Jensen2004,Lundbak2006}, for a given integral membrane protein, via systematic changes in the lipid acyl-chain length \cite{Perozo2002,Rawicz2000}. Furthermore, $U$ can also be modified experimentally through repositioning of amphipathic protein residues \cite{Draheim2006}. The general analytic expression of the MPPN energy and the corresponding \textcolor{black}{MPPN self-assembly diagram} obtained here show that pronounced protein-induced lipid bilayer thickness deformations favor highly symmetric and uniform MPPN shapes. Our results suggest strategies for producing highly symmetric and uniform MPPNs in experiments, and may thus help to optimize MPPN shape for structural studies of membrane proteins and targeted drug delivery.

\acknowledgments

This work was supported by NSF award numbers DMR-1554716 and DMR-1206332, an Alfred P. Sloan Research Fellowship in Physics, the James H. Zumberge Faculty Research and Innovation Fund at the University of Southern California, and the USC Center for High-Performance Computing. We also acknowledge support through the Kavli Institute for Theoretical Physics, Santa Barbara, via NSF award number PHY-1125915. We thank W.~S. Klug, R. Phillips, D.~C. Rees, M.~H.~B. Stowell, and H. Yin for helpful comments.


\includepdf[pages=1-]{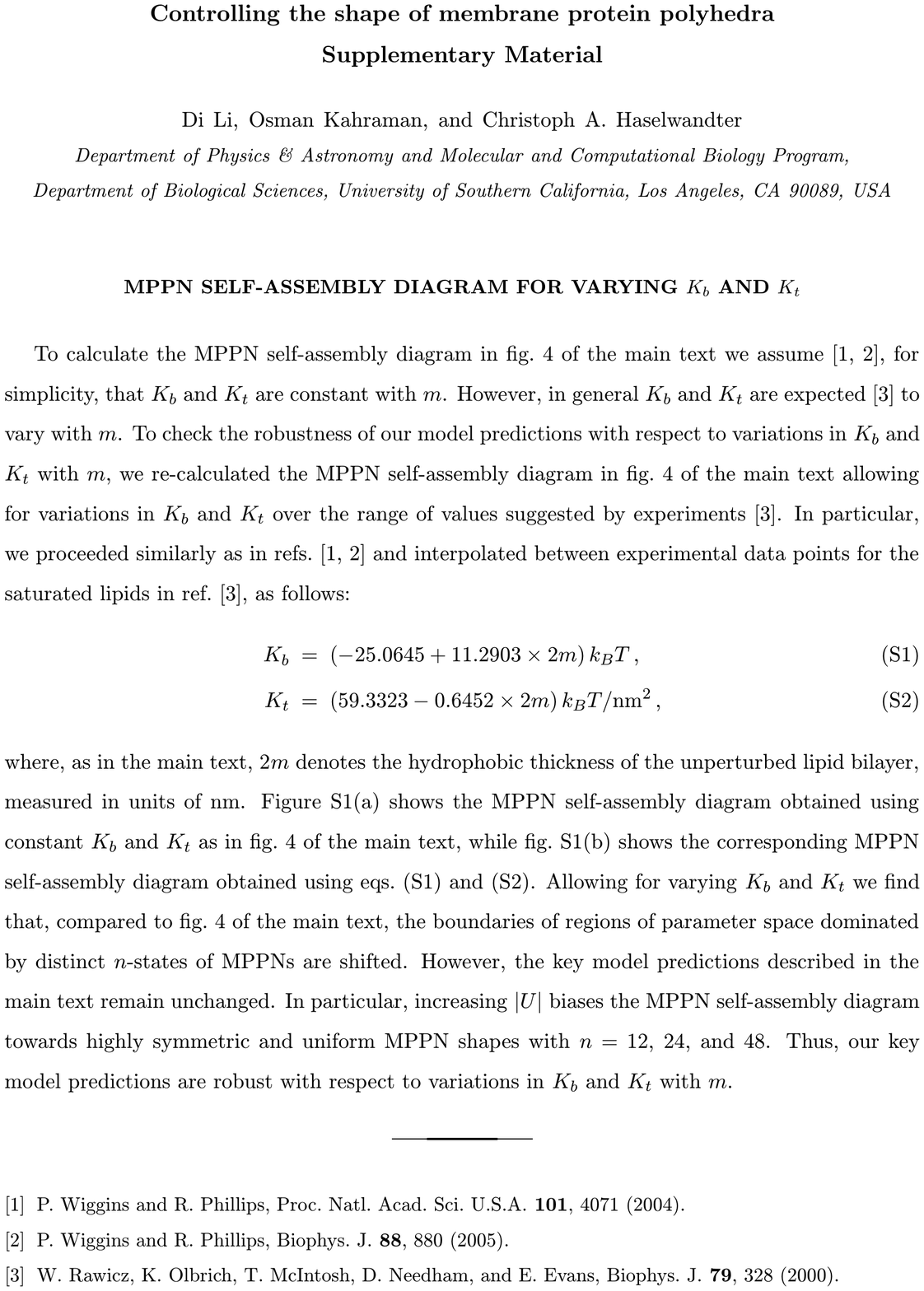}

\end{document}